\documentclass[11pt,a4paper]{article}
\usepackage{jheppub}

\usepackage{xcolor}
\usepackage{slashed}
\usepackage{dcolumn}
\usepackage{verbatim}
\usepackage{float}
\usepackage{multirow}
\usepackage{xspace}
\usepackage[normalem]{ulem}
\usepackage{url}
\usepackage{subfigure}
\usepackage{anyfontsize}
\usepackage{t1enc}

\definecolor{aquamarine}{rgb}{0.2,0.7,0.6}
\definecolor{cerulean}{RGB}{0,166,214}
\definecolor{hypershade}{rgb}{0.3,0.3,0.8}
\definecolor{subtlered}{rgb}{0.8,0.3,0.3}

\newcommand{\beq}{\begin{equation}}
\newcommand{\eeq}{\end{equation}}
\newcommand{\bea}{\begin{eqnarray}}
\newcommand{\eea}{\end{eqnarray}}

\definecolor{rosy}{RGB}{230,235,252}
\definecolor{myframetitle}{RGB}{90,89,170}
\definecolor{myblocktitle}{RGB}{140,185,249}
\definecolor{mytitle}{RGB}{10,80,26}

\definecolor{darkgreen}{RGB}{27,130,45}
\definecolor{darkblue}{rgb}{0,0,0.3}
\definecolor{darkred}{rgb}{0.7,0,0}

\definecolor{light gray}{RGB}{220,220,220}
\definecolor{dark purple}{RGB}{108,0,217}
\definecolor{pink}{RGB}{190,20,100}
\definecolor{orang}{RGB}{193,63,0}
\definecolor{green}{RGB}{11,98,17}
\definecolor{darkpink}{RGB}{153,0,76}
\definecolor{bluegreen}{RGB}{0,102,102}
\definecolor{greenlagan}{RGB}{0,102,0}
\definecolor{redgreen}{RGB}{102,102,0}
\definecolor{Redgreen}{RGB}{153,76,0}
\definecolor{vividviolet}{rgb}{0.62, 0.0, 1.0}
\definecolor{amaranth}{rgb}{0.9, 0.17, 0.31}
\definecolor{palatinateblue}{rgb}{0.15, 0.23, 0.89}
\definecolor{brightpink}{rgb}{1.0, 0.0, 0.5}
\definecolor{cornflowerblue}{rgb}{0.39, 0.58, 0.93}
\definecolor{deepcarminepink}{rgb}{0.94, 0.19, 0.22}
\definecolor{radicalred}{rgb}{1.0, 0.21, 0.37}

\def\rhox{\rho_\chi}
\def\vx{v_\chi}
\def\mx{m_\chi}
\def\mdm{m_\chi}

\def\Mdet{M_{\rm fid}}
\def\texp{t_{\rm exp}}
\def\Rdet{R_{\rm fid}}
\def\Erec{E_{\rm R}}
\def\nT{n_{\rm T}}

\def\mT{m_{\rm T}}
\def\muTx{\mu_{\rm T\chi}}

\def\tev{{\rm TeV}}
\def\gev{\mathrm{GeV}}

\def\sigmaTx{\sigma_{\rm T\chi}}

\def\sigmaTxeff{\sigma_{\rm T\chi}^{\rm eff}}

\def\Nhit{N^{\rm exp}_{\rm hit}}

\def\NBG{N_{\rm B}}

\def\NexpCL{N_{\rm exp}^{\rm 90CL}}

\allowdisplaybreaks

\title{Direct detection of electromagnetically interacting ultraheavy dark matter}

\author[a]{Jason Kumar,}
\author[b]{Biprajit Mondal,}
\author[b]{Girish Muralidhara,}
\author[b]{Nirmal Raj}

\affiliation[a]{Department of Physics \& Astronomy, University of Hawai'i, Honolulu, HI 96822, USA}
\affiliation[b]{Centre for High Energy Physics, Indian Institute of Science, C. V. Raman Avenue, Bengaluru 560012, India}

\emailAdd{jkumar@hawaii.edu}
\emailAdd{biprajitm@iisc.ac.in}
\emailAdd{nraj@iisc.ac.in}

\abstract{At a level too faint for astronomy, particle dark matter may interact with Standard Model states via the photon.
We derive limits from direct detection experiments on photon-mediated nuclear interactions up to operator dimension-6, viz., via a millicharge, charge radius, electric and magnetic dipole moment, and anapole moment, up to dark matter masses $\lesssim 10^{17}$~GeV,
where detectors become flux-limited.
We derive constraints for XENON1T, XENONnT, LZ, PANDAX-II, PANDAX-4T, DarkSide-50, DEAP-3600, PICO-60, and CDMS-II,
and estimate future sensitivities for the multi-deca-tonne scale experiments DarkSide-20k, DARWIN/XLZD, PANDAX-xT, and Argo.
Special attention is paid to millicharged dark matter, for which we constrain new parameter space by
deriving the ceiling on sensitivity to its electric charge, arising from the nuclear-vs-electron recoil discriminants used in liquid argon and bubble chamber detectors.
This result goes beyond ceilings previously identified in liquid xenon and semiconductor detectors from electron recoil vetoes.
In particular, the ceiling from PICO-60 closes a large window between direct detection and neutrino experiment limits for dark matter masses below $10^{12}$~GeV.}

\keywords{dark matter, direct detection, millicharged dark matter, electromagnetic interactions, ultraheavy dark matter}

\begin{document}

\maketitle

%%%%
\section{Introduction}
%%%%

``Dark'' matter, though invisible on astrophysical scales, may have particle interactions with electromagnetism~\cite{EMDM:Pospelov:2000bq,EMDM:Sigurdson:2004zp,EMDM:ChuzhoyKolb:2008zy,EMDM:Masso:2009mu,EMDM:Fitzpatrick:2010br,EMDM:Banks:2010eh,EMDM:Barger:2010gv,EMDM:McDermott:2010pa,EMDM:Fortin:2011hv,EMDM:Ho:2012bg,EMDM:DelNobile:2014eta,EMDM:Hambye:2021xvd,EMDM:Tomar:2022ofh,EMDM:Bose:2023yll,EMDM:Ibarra:2024mpq}.
In this paper we present up-do-date constraints from underground direct searches on ultraheavy dark matter (DM) that interacts with photons via electric charge, charge radius, electric and magnetic dipole moments, and the anapole moment.
Electrically charged -- ``millicharged'' -- DM has been extensively studied, with focus on sub-nuclear masses~\cite{EMDM:Ibarra:2024mpq,earthboundMCrelics:Pospelov:2020ktu,terradensity:BerlinLiuPospHari:2023zpn,Iles:2024zka}.
We will focus on the range $m_\chi \gg \tev$ for three reasons.
(i) By recasting direct searches we wish to present the latest limits, which go all the way up to $10^{16}$~GeV.
We identify the entire extent of regions constrained in cross section vs.~mass space from single-scatter searches, and clarify why a recast such as ours cannot be done with multi-scatter searches.
(ii) We carefully identify the cross section ceilings.
In Ref.~\cite{champCRs:Dunsky:2018mqs} it was recognized that the ceilings in liquid xenon and semiconductor detectors came from a veto on excess electron recoils (ERs),
since millicharged DM scatters on both nuclei and electrons.
Here we identify the ceiling for all current and imminent experiments.
In particular, we estimate for the first time ER-veto ceilings at liquid argon detectors and show that such a ceiling does not exist for bubble chamber detectors.
This allows us to constrain hitherto-unbounded parameter space.
(iii) Galactic magnetic fields could expel millicharged DM from the disk and suppress their flux at terrestrial experiments~\cite{EMDM:ChuzhoyKolb:2008zy} (although see Ref.~\cite{EMDM:PerriFarrarBFieldsReal:2025tvq}), but this effect goes away at high DM masses, so that direct searches are at play again.
This is because the disk evacuation constraints on the millicharge scale as the DM mass $m_\chi$, whereas direct detection limits scale more gently as $m^{1/2}_\chi$.

The other electromagnetic operators we consider do not enjoy subtleties (ii) and (iii); nonetheless, we estimate the full extent of their limits for reason (i).
A special feature we consider is single-scatter ceilings as treated in Ref.~\cite{RajNuRoof:2024guv}.
A number of other models of ultraheavy DM are also constrained at direct detection experiments~\cite{Models:Chung1998wimpzillas,Models:kolb1998wimpzillas,Models:Harigaya2016:GUTzillas,Models:ElectroweakBalls,Models:BNL:DarkBaryonGeVMediator,Models:ColoredDM,Bramante:2019yss,Bramante:2018tos,Models:EWSymMonopoles:Bai:2020ttp,Models:nucleiHardy:2014mqa,Models:nucleiHardy:2015boa,Models:nucleiMonroe:2016hic,Models:Nuggets,Models:PlanckScaleBHRelicsBaiOrlofsky2019,Models:PlanckScaleBHRelicsSantaCruz,Aggarwal:2024ngx,snowmass:Carney:2022gse}.

This paper is organized as follows.
In Sec.~\ref{sec:setup} we provide the interaction Lagrangian and set up the direct detection formalism, including a description of ceilings on the millicharge of DM from electron recoil vetoes.
In Sec.~\ref{sec:results} we display and discuss our results.
In Sec.~\ref{sec:concs} we conclude and discuss future avenues.
In the appendix we collect expressions for scattering cross sections in terms of nuclear matrix elements and form factors.

%%%%%
\section{Set-up}
\label{sec:setup}
%%%%%

To study the full extent of electromagnetic interaction scenarios involving effective operators up to dimension-6, we will take DM to be a spin-1/2 Dirac state.
(Some operators vanish for spin-0 and Majorana DM~\cite{DelNobile:2021wmp}.)
The effective Lagrangian for DM-photon interactions is then given by
%%%%%
\begin{eqnarray}
\nonumber \mathcal{L}_{\rm int} = && Q_\chi e \bar \chi \gamma^\mu \chi  A_\mu
+ \frac{\mu_\chi}{2} \bar \chi \sigma^{\mu \nu} \chi F_{\mu \nu}
+ i \frac{d_\chi}{2} \bar \chi \sigma^{\mu \nu} \gamma^5 \chi F_{\mu \nu} \\
+&& \mathcal{A}_\chi \bar \chi \gamma^\mu \gamma^5 \chi \partial^\nu F_{\mu \nu}
+ b_\chi e\bar \chi \gamma^\mu \chi \partial^\nu F_{\mu \nu}~,
\label{eq:Lint}
\end{eqnarray}
%%%%%
where the five terms correspond to millicharge ($Q_\chi$),
magnetic dipole moment ($\mu_\chi$),
electric dipole moment ($d_\chi$),
anapole moment ($\mathcal{A}_\chi$), and
charge radius ($b_\chi$) interactions.
We will study each scenario separately by turning off the other couplings.

For a recoil energy range $\{ E_{\rm R,min}, E_{\rm R,max}\}$ on a nuclear target $T$ detected with efficiency $\epsilon_{\rm NR}$ in a fiducial volume of radius $R_{\rm fid}$ exposed over a time $t_{\rm exp}$, the expected number of single-scatter events is~\cite{RajNuRoof:2024guv}
%%%
\begin{equation}
N_{\rm ev}^{\rm SS} = \epsilon_{\rm NR} \  p_{\rm hit}(1, \Nhit) \  \Phi_{\rm int}~,
\label{eq:NevSSmaster}
\end{equation}
%%%
where the integrated flux of DM of mass $\mx$ is\footnote{For millicharged dark matter the flux may be affected by supernova shocks and diffusion into the Galactic disk by magnetic fields, but in the $>$ TeV mass range we consider these have negligible effect~\cite{champCRs:Dunsky:2018mqs}.}
%%%%
\begin{equation}
\Phi_{\rm int} = \bigg(\frac{\rhox}{\mx}\bigg) \pi \Rdet^2 \bar v \texp
\label{eq:integfux}
\end{equation}
%%%%
with density $\rhox = 0.3~$GeV/cm$^3$ and average DM speed $\bar v = 270$~km/s.
For nuclear target number density $\nT$ and average transit length $L_{\rm ave}$,
the expected number of DM-nucleus scatters per transit -- the multiplicity -- is
%%%
\begin{equation}
\Nhit = \sigmaTxeff \nT L_{\rm ave}~,
\label{eq:multiplicity}
\end{equation}
%%%
and we take $p_{\rm hit}(1, \Nhit)$ to follow the Poisson probability for obtaining a single scatter when $\Nhit$ are expected.
As in Ref.~\cite{RajNuRoof:2024guv}, to determine $L_{\rm ave}$ we assume detectors have a spherical geometry, an approximation that reproduces experimental results closely.
For a DM-target reduced mass $\muTx$, we can write an effective cross section
%%%
\begin{equation}
\sigmaTxeff = \frac{1}{\bar v} \int^{E_{\rm R, max}}_{E_{\rm R, min}}  d\Erec \bigg\langle \frac{d\sigmaTx}{d\Erec} \bigg\rangle \int_{v_{\rm min}(\Erec)}^{v_{\rm esc}}  d^3 v~v f_{\rm lab}(\Vec{v})~,
\label{eq:sigmaTxeff}
\end{equation}
%%%
where
$\langle ... \rangle$ denotes summing over relevant isotopes weighted by their number fraction,
$v_{\rm min} (\Erec) = \sqrt{\mT \Erec/2\muTx^2}$ is the minimum DM speed required to produce a nuclear recoil of energy $\Erec$,
and
$v_{\rm esc} = 544$~km/s is the Galactic escape speed.
As shown in Ref.~\cite{RajNuRoof:2024guv}, for small cross sections where $\Nhit \ll 1$, Eq.~\eqref{eq:NevSSmaster} reduces to the familiar expression seen in e.g., Refs.~\cite{Lewin:1995rx,DelNobile:2021wmp} for the optically-thin limit.
In general, $d\sigmaTx/d\Erec$ could contain a sum of spin-independent and spin-dependent nuclear responses and form factors, depending on the Galilean-invariant operators involved in the non-relativistic limit; their expressions for our five scenarios are given in Appendix~\ref{sec:CrossSections}.

%%%%%%%%
\begin{table}[t]
    \small
    \centering
    \resizebox{1.15\textwidth}{!}{%
    \begin{tabular}{|p{34mm}|p{37mm}|p{23mm}|p{14mm}|p{5mm}|p{40mm}|p{28mm}|}
    \hline
   &  & $\Mdet \times \texp$  & &  & & \\
  target  & detector  & (ton $\times$ yr) & $\Erec$ (keV) & $\epsilon_{\rm NR}$ &  ($N_{\rm B}\pm\sigma_B$, $N_{\rm obs}, N_{\rm exp}^{\rm 90CL})$ & $Q_\chi$ ceiling  \\
     \hline
       &  \cite{DARWIN:Macolino:2020uqq}~{\bf DARWIN/XLZD} & \mbox{40 $\times$ 5} & [5, 35] & 0.50 & (4.1, 4.1, 4.0) & \\
       & \cite{PandaX-xT:2024oxq}~{\bf PANDAX-xT} & \mbox{34.2 $\times$ 5.85} & [4, 35] & 0.50 & (48$\pm$6.9, 48, 11.4) & \\
    xenon,   & \cite{XENON1T:MS:2023iku}~XENON1T & \mbox{1.3 $\times$ 0.76} & [10, 40] & 0.80 & (7.4$\pm$0.6, 14, 12.8) & \\
  \scriptsize{2.94 g/cm$^3$}   & \cite{XENONnT:SS:2025vwd}~XENONnT & \mbox{4.18 $\times$ 0.77} & [10, 40] & 0.80 & ($-$,$-$, 14.2) & ER veto \\
  \scriptsize{$^{128}$1.9\%, $^{129}_{\rm SD}26.4\%$, $^{130}$4.1\%}  & \cite{LZ:SS:2024zvo}~LZ & \mbox{5.5 $\times$ 0.77} & [5, 50] & 0.90 & ($-$,$-$, 21.2) & \\
   \scriptsize{ $^{131}_{\rm SD}$21.2\%, $^{132}$26.9\%,}    & \cite{PandaX-II:SS:2020oim}~PANDAX-II & \mbox{0.33 $\times$ 1.1} & [10, 30] & 0.85 & (40.3$\pm$3.1, 38, 7.8) & \\
   \scriptsize{ $^{134}$10.4\%, $^{136}$8.9\%}    & \cite{PandaX-4T:SS:2021bab}~PANDAX-4T & \mbox{2.67 $\times$ 0.24} & [30, 90] & 0.75 & (9.8$\pm$0.6, 6, 0.8) & \\
         \hline
          & \cite{DarkSide-20k:2017zyg}~{\bf DarkSide-20k} & \mbox{20 $\times$ 10} & [30, 200] & 0.90 & (3.2, 3.2, 3.7) & \\
     & \cite{ARGO:2018}~{\bf Argo} & \mbox{300 $\times$ 10} & [55, 100] & 0.90 & (15.6$\pm$5, 15.6, 6.4) & \\
    argon,     & \cite{DarkSide:SS:2018kuk}~DarkSide-50 & \mbox{0.031 $\times$ 1.46} & [80, 200] & 0.70 & (0, 0, 2.3) & early veto system \\
   \scriptsize{1.40 g/cm$^3$}      & \cite{DEAP:SS:2019yzn}~DEAP-3600 & \mbox{0.824 $\times$ 0.63} & [70, 100] & 0.24 & (0, 0, 2.3) & for ER vetos \\
     \hline
  C$_3$F$_8$,~\scriptsize{8.17 g/cm$^3$} & \cite{PICO60:2019vsc}~PICO-60 & \mbox{0.049 $\times$ 0.08} & [4, 15] & 0.96 & (0, 3, 6.7) & no ER veto\\
        \hline
        Ge,~\scriptsize{5.32 g/cm$^3$} & \cite{CDMS-II:2009ktb}~CDMS-II & \mbox{0.0032 $\times$ 0.52} & [10, 100] & 0.25 & (0, 2, 5.3) & ionization yield \\
        \hline
    \end{tabular}%
   }
    \caption{
       Experiments (with proposed forthcoming ones in {\bf bold}) for which limits are shown in Figs.~\ref{fig:DDlimitsMCP} and \ref{fig:DDlimitsEDMMDMCRADM}.
     Given here are the detector fiducial mass $\Mdet$, live time $\texp$, signal efficiency/nuclear recoil acceptance $\epsilon_{\rm NR}$, background estimate $\NBG$ with uncertainty $\sigma_B$, observed event count $N_{\rm obs}$, and the 90\% C.L. limit $\NexpCL$ as estimated in Ref.~\cite{RajNuRoof:2024guv}.
     Noble liquid densities and isotope abundances are provided in the first column, with the subscript ``SD'' indicating xenon isotopes that have spin-dependent interactions.
     The expected event count for XENONnT and LZ are obtained by scaling up those in Ref.~\cite{RajNuRoof:2024guv} by the exposure here, which closely matches the experimental exclusion curves.
     The table reflects the fact that CDMS-II used 14 out of their 19 Ge detectors, and none of the Si detectors; we display limits for CDMS-II as opposed to its successor Super-CDMS~\cite{SuperCDMS:2014cds} due to its larger exposure.
     The last column gives the criterion for the lower limit (ceiling) on the dark matter millicharge $Q_\chi$ as detailed in Sec.~\ref{subsec:ERveto}.}
    \label{tab:detectordetails}
\end{table}

%%%

%%%%%
\begin{figure*}[t]
    \centering
     \includegraphics[width=.45\textwidth]{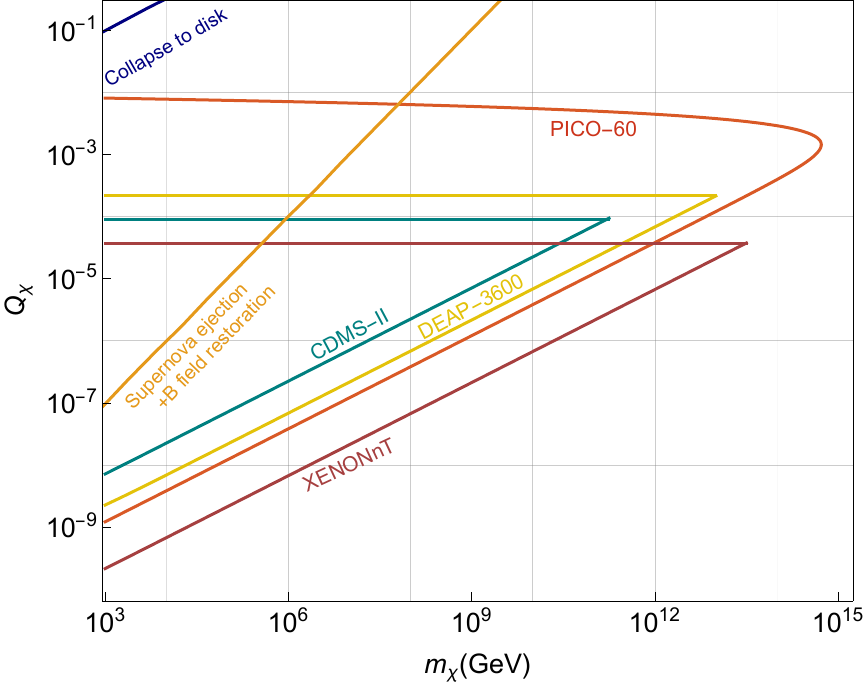}
     \includegraphics[width=.45\textwidth]{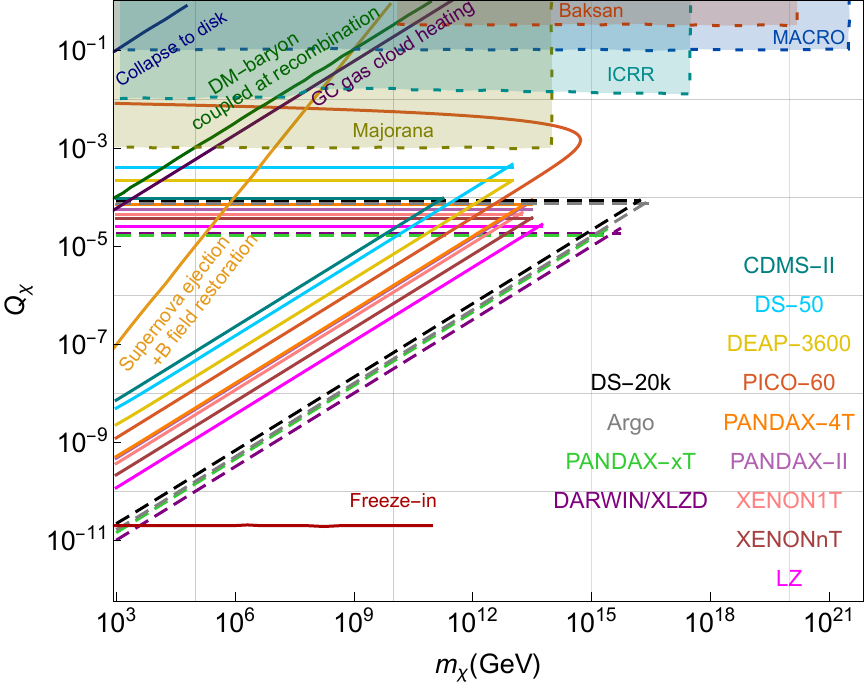}
    \caption{90\% C.L. limits (solid curves) and future sensitivities (dashed curves) from single scatter- and nuclear recoil-based direct detection searches on the electric charge of dark matter as a function of its mass, obtained using Table~\ref{tab:detectordetails} and the statistical method in Ref.~\cite{RajNuRoof:2024guv}.
    On the left are limits from four different experiments that typify how the ceiling on $Q_\chi$ is obtained for their kind, and on the right are all experiments considered in this work.
    Also shown are parameters that produce the dark matter abundance via freeze-in as reproduced from Ref.~\cite{champCRs:Dunsky:2018mqs}; and regions where the millicharged DM collapses into the Galactic disk, is ejected by supernova shocks while restocked by Galactic $B$ fields, as well as limits from DM-baryon coupling during recombination, from other terrestrial detectors as estimated in Ref.~\cite{champCRs:Dunsky:2018mqs}, and from heating of gas clouds at the Galactic Center.
    See Sec.~\ref{sec:results} for further details.}
\label{fig:DDlimitsMCP}
\end{figure*}
%%%%

%%%%%
\begin{figure*}[t]
    \centering
     \includegraphics[width=.45\textwidth]{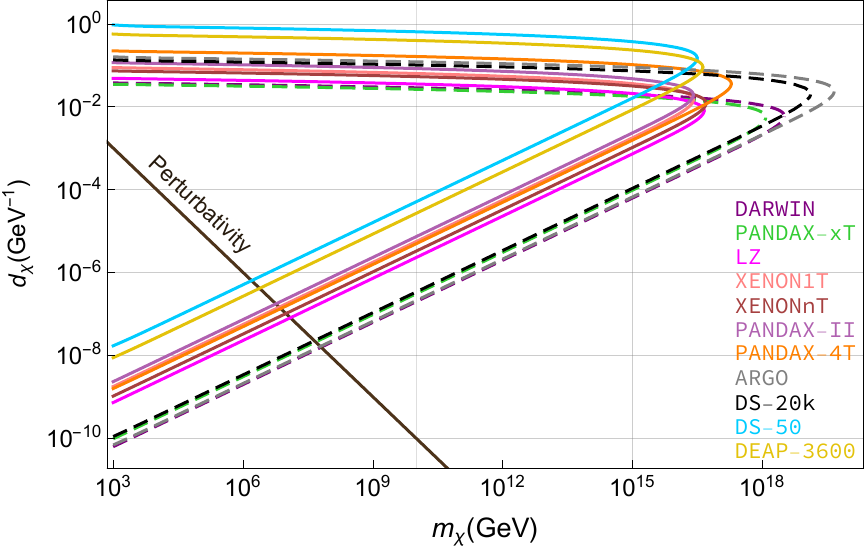}
    \includegraphics[width=.45\textwidth]{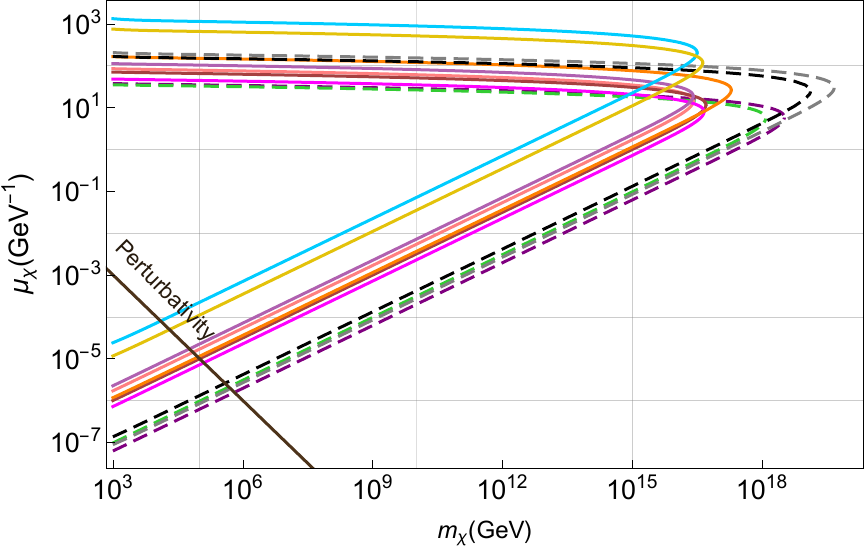} \\
    \includegraphics[width=.45\textwidth]{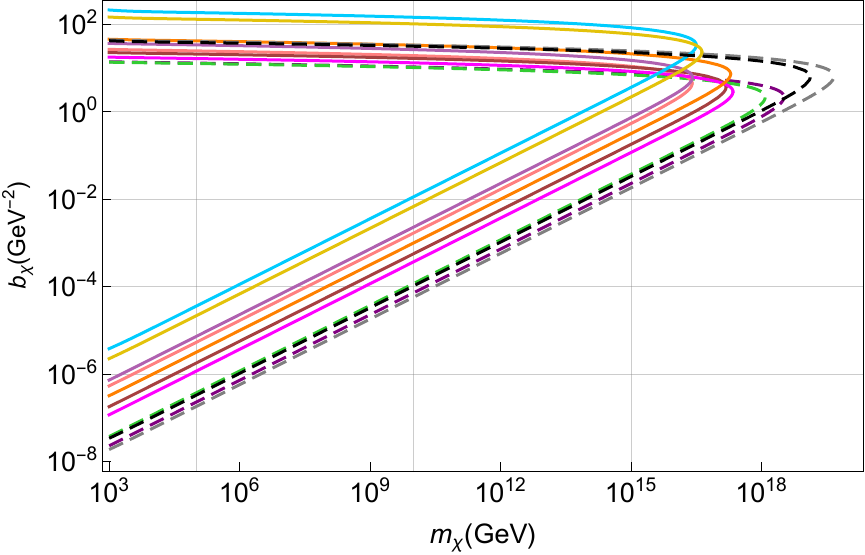}
    \includegraphics[width=.45\textwidth]{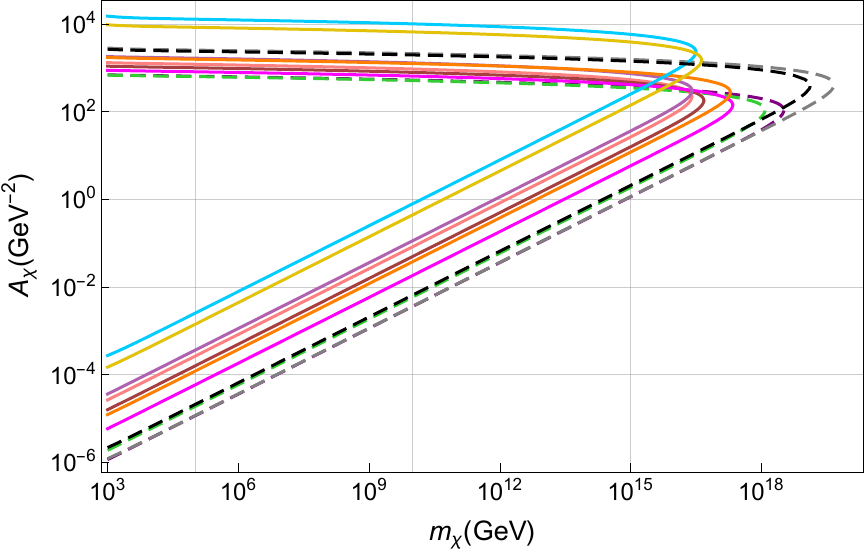}
    \caption{90\% C.L. limits and sensitivities as a function of dark matter mass for electric dipole moment (top left), magnetic dipole moment (top right), charge radius (bottom left), and anapole moment (bottom right) of dark matter.
    The plotting scheme is the same as in Fig.~\ref{fig:DDlimitsMCP}.
    In the regions $d_\chi < e/m_\chi$ and $\mu_\chi < e/\mx$ we expect a simple perturbative field configuration.
    See Sec.~\ref{sec:results} for further details.}
\label{fig:DDlimitsEDMMDMCRADM}
\end{figure*}
%%%%

%%%%
\subsection{Ceilings from electron recoil veto for milli-charged dark matter}
\label{subsec:ERveto}
%%%%

Milli-charged DM scattering on SM states, proceeding via the massless photon mediator, has a differential cross section $d\sigma/d\Erec \propto Z^2 Q_\chi^2/q^4$, favoring scatters with small momentum transfer $q$.
For nuclear scatters, the minimum detectable $q = \sqrt{2 \mT \Erec^{\rm th}}$ for threshold energy $\Erec^{\rm th} = \mathcal{O}$(keV) is $\mathcal{O}$(10 MeV), whereas for atomic electron scatters it is $\sim E_{\rm bind}/\vx$, i.e. $\mathcal{O}$(10 keV)~\cite{Catena:2019gfa}.
Comparing the cross sections, we then see that milli-charged DM-electron interactions dominate over DM-nucleus interactions in the detector during DM transit.
This implies that as we increase $Q_\chi$ beyond some point, electron recoil (ER) signals -- that are typically vetoed -- would limit searches for nuclear recoil (NR) signals.
Hence one generally expects a ceiling in $Q_\chi$-$\mx$ space from ER vetoes when recasting single-scatter direct searches to millicharged DM.
Such a ceiling doesn't apply to the other operators we consider here because the associated differential cross sections are not as strongly-peaked at
small $q$, so the difference between their NR and ER cross sections is not as large.
In Table~\ref{tab:detectordetails} we have mentioned in brief the critera we use to set our ceilings on $Q_\chi$, which we now describe.

As discussed in Ref.~\cite{NISTStoppingPower}, the stopping power ($dE/dx$) of millicharged DM by electrons in our velocity regime is given by $Q_\chi^2$ times the proton stopping power, taken from Ref.~\cite{NISTStoppingPower}.
We will use these to compute ionization losses in various detector materials.
In Ref.~\cite{champCRs:Dunsky:2018mqs}, the ceiling on $Q_\chi$ was identified for two types of detectors.
For liquid xenon (LXe) detectors, and specifically for XENON1T, the ER veto was taken as the ceiling criterion.
A conservative limit was estimated by requiring effectively zero electron recoils: the ionization energy loss of DM through 1~m of LXe was restricted to $<$~10~eV.
We will impose a similar condition by restricting the ionization energy loss to $<$~10~eV over a distance $L_{\rm ave}$ for each of the LXe detectors we consider.
Ref.~\cite{champCRs:Dunsky:2018mqs} also estimated the ceiling at CDMS-II by requiring the DM ionization energy loss through 1 cm of Ge to be $<$~3~keV.
This is to satisfy the condition that the so-called ionization yield -- the ratio of ionization to NR energy (as defined by CDMS~\cite{CDMS-II:2009ktb}) -- is $<$ 0.3.

For other detector types, other considerations apply for estimating the ceiling, which we now describe.
In dual-phase argon detectors, NRs and ERs are distinguished via pulse shape discrimination (PSD).
Hence in addition to the photoelectron count, PSD variables such as $F_{\rm prompt}$, $f_{90}$ and $f_{200}$ provide an additional handle to identify signal regions~\cite{DEAP:SS:2019yzn,DarkSide:SS:2018kuk,DarkSide-20k:2017zyg}.
In principle, this could help in vetoing ERs from millicharged DM at large $Q_\chi$ (a luxury not enjoyed by xenon TPC detectors, which only use photoelectron counts to define signal regions).
However, in practice, when an NR is picked up an ``early veto system'' looks for ERs in the 2.5~$\mu$s preceding the event at DEAP-3600; if 4 or more ERs are detected, the event is vetoed~\cite{swprivcorr}.
As this is also the timescale of DM transit through the detector, and since (as argued here) the DM scattering cross section for ERs is greater than NRs, we can expect the early veto system to come into play for values of $Q_\chi$ that make the detector optically thick in the DM-electron scattering channel.
Accordingly, adopting this criterion for liquid argon detectors, we mark the NR single-scatter ceiling as the $Q_\chi$ for which the ionization loss is 40 eV, corresponding to 4 ERs per transit.
(This is a somewhat conservative criterion as the ERs may also occur {\em after} the NR, in which case the early veto system wouldn't pick it up.)
The ceiling can be pushed higher in future dedicated searches for millicharged DM by redoing offline cuts such that the early veto is not triggered.

In bubble chamber detectors, a superheated liquid is prepared under the right pressure conditions such that a stable bubble is nucleated if energy is deposited above some critical value within a critical volume.
This allows for clear distinction between NRs and ERs that respectively produce large and small ionization densities~\cite{bubchambdiscrim:Aubin:2008qx} (as arising respectively from their $\mathcal{O}$(10 keV) and $\mathcal{O}$(10 eV) energy depositions).
And in fact, in the context of bubble chamber detectors it is the discrimination between NRs and alpha particles that is relevant.
This implies the single-scatter NR ceiling in detectors such as PICO is not determined by an ER veto criterion, but by the values of $Q_\chi$ that set the boundary between single-scatter and multi-scatter signals.
Thus we will simply use the formalism of Ref.~\cite{RajNuRoof:2024guv}, applied to the set-up in Eq.~\eqref{eq:NevSSmaster}.
This lets us set limits at high DM masses on large values of $Q_\chi$ that were missed in the previous literature, allowing us to close a parametric window.

%%%%%
\section{Results}
\label{sec:results}
%%%%%

We show our millicharged limits and reaches in the $Q_\chi$-$\mx$ plane in Fig.~\ref{fig:DDlimitsMCP}. In the left panel are limits from XENONnT, DEAP-3600, CDMS-II, and PICO, derived from the event rate in Sec.~\ref{sec:setup} and using the experimental parameters in Table~I of Ref.~\cite{RajNuRoof:2024guv}, and considerations of the ER veto ceiling in Sec.~\ref{subsec:ERveto}.
This panel highlights the comparison across ceilings at different kinds of detectors.
In the right panel we show limits for all the experiments in Table~\ref{tab:detectordetails}; the curves for XENON1T and CDMS-II closely match those in Ref.~\cite{champCRs:Dunsky:2018mqs}.
In both panels are lines that indicate regions where millicharged DM collapses into the Galactic disk during formation ($Q_\chi > \sqrt{\mx/10^5~{\rm GeV}}$) and where it is ejected by supernova shocks though replenished by diffusion through inhomogeneous magnetic fields ($\sqrt{\mx/10^5~{\rm GeV}} > Q_\chi > \mx/10^{10}~{\rm GeV}$), as derived in Ref.~\cite{champCRs:Dunsky:2018mqs}.
Also shown in the right panel are limits from the heating of Galactic Center gas clouds\footnote{These limits are
thought to be valid as millicharged DM-deflecting magnetic fields in the Galactic Center are oriented not parallel to the disk, but toward the halo edge~\cite{gascloudPRLMCP:Bhoonah:2018wmw}.} and limits derived in Ref.~\cite{champCRs:Dunsky:2018mqs} from MACRO, ICRR, MAJORANA and Baksan data.

Both panels highlight a central result of our study: the limits from the bubble chamber-based PICO-60 extend to two orders higher in millicharge than other experiments and one order greater in DM mass due to there being no ceiling from electron recoil vetoes.
Due to the somewhat relaxed criterion for the ceiling in argon detectors (from the early veto system) than in xenon detectors, we see that DEAP-3600 and DarkSide-50 set limits on $Q_\chi$ up to $\mathcal{O}(10^{-4})$, which is higher than xenon ceilings.
The right panel also highlights that future experiments would probe parameter space where millicharged DM is produced via cosmological freeze-in (as estimated in Ref.~\cite{champCRs:Dunsky:2018mqs}).
Since the regions we limit are necessarily where ERs in detector transit are negligible, and since we only consider large DM masses with large kinetic energies, we do not expect that ceilings on $Q_\chi$ from DM stopping in the Earth's crust via scattering on electrons would place stronger limits than our ER veto-based ceilings.
For $\mathcal{O}(1)$ ERs in a meter-scale detector, we expect $\mathcal{O}(10^3)$ ERs in the km-scale crust overburden; as the fractional energy loss per scatter $m_e/\mx \sim 10^{-6}$, for $\mx >$~TeV DM scatters in the crust would scarcely distort its velocity distribution.
By this reasoning, we also expect both ERs and NRs in the overburden to have a negligible effect on the DM velocity distribution in the case of PICO-60, which does not have an ER veto-based ceiling, for $\mdm \gtrsim 10^5$~GeV where we constrain new parameter space.
A careful study of the Earth overburden for millicharged DM across all DM masses would nonetheless be an interesting investigation.

In Fig.~\ref{fig:DDlimitsEDMMDMCRADM} we show limits as a function of the DM mass on the magnetic dipole, electric dipole, anapole moments, and the charge radius.
As there is no ER veto ceiling for these scenarios we estimate the single-scatter ceiling simply from Eq.~\eqref{eq:NevSSmaster}.
As can be seen from the Poisson probability in Eq.~\eqref{eq:NevSSmaster} and discussed at length in Ref.~\cite{RajNuRoof:2024guv}, this ceiling is logarithmically sensitive to $\mx$ and to the exposure time; the limit in cross section scales with the fiducial mass as $M_{\rm fid}^{-1/3} \log M_{\rm fid}^{2/3}$.
To our knowledge, this is the first time the full extent of limits from single-scatter direct detection on these operators is being displayed.
We also show perturbative limits of $d_\chi \leq e/\mx$ and $\mu_\chi \leq e/\mx$; beyond these regions we assume that DM has some non-perturbative field configuration (see, e.g., Ref.~\cite{Sigurdson:2004zp}).
In particular, we note that the electric and magnetic dipole moment
ceilings of the experiments
considered extend all the way into the non-perturbative regime.
The top left regions of these figures near our ceilings may be constrained by experiments in rockets, balloons and on the Earth's surface~\cite{Sigurdson:2004zp,Li:2022idr}; we leave to future a careful recast of these searches.
In both Fig.~\ref{fig:DDlimitsMCP} and Fig.~\ref{fig:DDlimitsEDMMDMCRADM} we show limits for $m_\chi \geq$ TeV as that is the range of our interest for reasons discussed in the Introduction -- to close new parameter space, to highlight our new ceilings, and in the case of millicharged DM, to operate in a region where the DM flux is certain.
Other limits exist at lower masses too, which the reader may find in the references in the Introduction.

%%%%%%
\section{Discussion}
\label{sec:concs}
%%%%%%

In this work we derived limits and sensitivities from direct searches of superheavy dark matter interacting electromagnetically.
We did this in the single-scatter (optically thin) regime, essentially by recasting WIMP searches.

A key finding of our work pertains to ceilings on the electric charge of DM.
Millicharged DM, interacting with both nuclei and electrons, is generally more likely to produce an ER above threshold than an NR.
This results in a ceiling to the sensitivity of experiments that can observe ERs, and which will reject almost all DM events as background for millicharge large
enough for the detector to be optically thick to ERs.
But as ERs leave no detectable signature in bubble chambers like PICO-60, no such ceiling applies to the detector type, leading to a sensitivity for PICO-60 two orders of magnitude higher than that of other underground DM direct searches (both current
and planned).
Indeed, PICO-60 closes the window between other DM direct detectors and neutrino experiments
(such as MAJORANA) for $m_\chi \lesssim 10^{12}~\gev$.

As for liquid xenon detectors, we had re-used the conservative criterion in Ref.~\cite{champCRs:Dunsky:2018mqs} of marking the ceiling where one ER is expected in the detector volume, but the ceiling can be pushed to higher values of $Q_\chi$ by a careful analysis of $S_1$ and $S_2$ pulses that would accommodate more DM-induced ERs.
Improvements are also possible by fitting to NR vs ER recoil spectra (see Appendix~\ref{sec:CrossSections}).
Similarly, we identified a conservative ceiling on $Q_\chi$ from liquid argon detectors by requiring not more than three ERs in the transit of DM.
This can be considerably relaxed by rewriting the early veto system in a targeted search for millicharged DM.

Direct searches for electron recoils would constrain our models.
For millicharged DM, Ref.~\cite{champCRs:Dunsky:2018mqs} recast XENON10's ER search with 0.04 kg-yr exposure (Ref.~\cite{Essig:2017kqs}) to set a rough limit of $Q^2_\chi/\mdm < 10^{-16}/$TeV for $\mx \gtrsim$~TeV; scaling this naively to XENONnT's 7.8 tonne-yr exposure ionization-only search in Ref.~\cite{XENON:2026qow} would result in a limit of $Q^2_\chi/\mdm < 5 \times 10^{-22}/$TeV, which is competitive with the upper limits in our Fig.~\ref{fig:DDlimitsMCP}.
A precise estimation of the latest limits would involve a careful treatment of atomic form factors, energy thresholds, and new criteria for ceilings, which we leave to a future study.
We also did not recast searches for multiscatter signals at large cross sections such as in Refs.~\cite{DEAP:MS:2021raj,PICO:MS:Broerman2022,XENON1T:MS:2023iku,LZ:MS:2024psa} as the cuts devised in many of these searches pick out nuclear scatters only, whereas we generically expect both nuclear and electron scatters, particularly in the case of millicharged DM (Sec.~\ref{subsec:ERveto}).
A detailed study of multiscatter signatures from large coupling in various detectors we leave to future work.

%%%
\section*{Acknowledgments}
%%%%

For collaboration in the early stages of this work we thank Shriti Saini.
We are also grateful to
Akash Kumar Saha and
Tracy Slatyer
for fruitful discussion, and
Shawn Westerdale
for reading the manuscript and offering helpful comments.
N.~R.~acknowledges support from the grant ANRF/ECRG/2024/000387/PMS.
J.~K.~is supported in part by DOE grant DE-SC0010504.
J.~K.~ is grateful to the Indian Institute of Science for its hospitality.

%%%%%%%%%%%%%%%%%
\appendix

%%%%%
\section{Cross sections}
%%%%%
\label{sec:CrossSections}

We provide here expressions for the differential cross sections for DM-nucleus scattering via the interactions in Eq.~\eqref{eq:Lint}.
Wherever derived elsewhere, these match with the literature e.g., Ref.~\cite{DelNobile:2021wmp}.
The form factors that appear below have lengthy expressions that we do not reproduce here, but they are listed in Refs.~\cite{Fitzpatrick:2012ix,AbdelKhaleq:2023ipt} to which we refer the reader.

For millicharged DM, and DM with charge radius and electric dipole moment,
%%%%%%
\begin{eqnarray}
\frac{d\sigmaTx}{d\Erec}\bigg|_{\rm MCP} &=& \frac{8\pi \mT}{v^2}\frac{1}{q^4}\alpha^2Q_\chi^2 Z^2 F^2(E_R)~,\\
\frac{d\sigmaTx}{d\Erec}\bigg|_{\rm CR} &=& \frac{2\pi}{9}\frac{\mT}{v^2}Z^2\alpha^2b_{\chi}^2F^2(E_R)~,\\
 \frac{d\sigmaTx}{d\Erec}\bigg|_{\rm EDM} &=& \frac{2 \mT}{v^2}\alpha d_\chi^2\frac{Z^2}{q^2}F^2(E_R)~,
\end{eqnarray}
%%%%%%
where $F(E_R)$ is the Helm form factor~\cite{Helm:1956zz}.
In principle, for millicharged DM there is also an atomic form factor that screens the nuclear charge, which effectively suppresses the cross section for
$q \lesssim 2\pi ({\rm angstrom})^{-1} \sim 10$~keV.
As our minimum detectable $q$ is $\mathcal{O}$(10 keV), we expect at most $\mathcal{O}(1)$ suppression and thus neglect the effect of this form factor.

For DM with magnetic dipole and anapole moments~\cite{DelNobile:2021wmp},
%%%%%
\begin{eqnarray}
\nonumber && \frac{d\sigmaTx}{d\Erec}\bigg|_{\rm MDM}=\frac{1}{2} \frac{\mT}{m_N^2}\frac{1}{v^2}\alpha \mu_{\chi}^2\Bigg[\Bigg(\frac{v^2}{\Erec}-\frac{\mx+2\mT}{2 \mx\mT}\Bigg)2\frac{m_N^2}{\mT}F_M^{(p,p)}(q^2)\\
\nonumber &&+ \ 4F_{\Delta}^{(p,p)}(q^2)-2\sum_Ng_NF_{\Sigma',\Delta}^{(N,p)}(q^2)+\frac{1}{4}\sum_{N,N'}g_Ng_{N'}F_{\Sigma'}^{(N,N')}(q^2)\Bigg]~,\\
\end{eqnarray}
%%%%%%
and
\begin{eqnarray}
\nonumber && \frac{d\sigmaTx}{d\Erec}\bigg|_{\rm ADM}=\frac{1}{2}\frac{\mT}{m_N^2}\frac{1}{v^2}q^2\alpha \mathcal{A}_{\chi}^{2}\Bigg[4m_N^2\Bigg(\frac{v^2}{q^2}-\frac{1}{4\mu_{\rm T}^2}\Bigg)F_M^{(p,p)}(q^2)\\
\nonumber &&   +4F_{\Delta}^{(p,p)}(q^2)-2\sum_Ng_NF_{\Sigma',\Delta}^{(N,p)}(q^2)+\frac{1}{4}\sum_{N,N'}g_Ng_{N'}F_{\Sigma'}^{(N,N')}(q^2)\Bigg]~,\\
\end{eqnarray}
%%%%%%
where $\mu_{\rm T}$ is the DM-nucleus reduced mass,
$F_M^{(p,p)}$ is the protonic magnetic monopole form factor,
$F_\Delta^{(p,p)}$ is the protonic longitudinal form factor,
$F_{{\Sigma'}}^{(N,N')}$ is the nucleonic transverse magnetic spin form factor, and
$F_{{\Sigma'},\Delta}^{(N,p)}$ are the nucleonic and protonic interference form factors, taken from Refs.~\cite{Fitzpatrick:2012ix,AbdelKhaleq:2023ipt}.
When taking the isotopic average in Eq.~\eqref{eq:sigmaTxeff}, the form factors $F_\Delta^{(p,p)}$, $F_{\Sigma',\Delta}^{(N,p)}$, and $F_{\Sigma'}^{(N,N')}$ do not vanish only for isotopes with odd $A$, while $F_M^{(p,p)}$ is present for all the isotopes.

%%%%
\bibliographystyle{JHEP}  % requires JHEP.bst in the same folder (see note below)
\bibliography{references}
%%%%

\end{document}